\documentclass[conference]{IEEEtran}
\IEEEoverridecommandlockouts
\usepackage{mathrsfs} % https://www.ctan.org/pkg/mathrsfs 

\usepackage{xcolor}

\ifCLASSINFOpdf
   \usepackage[pdftex]{graphicx}

\else
   \usepackage[dvips]{graphicx}
\fi

\usepackage{amsmath,amssymb}
\usepackage{array}
\usepackage{fixltx2e}
\usepackage{stfloats}

\begin{document}

\makeatletter
\newcommand{\vast}{\bBigg@{4}}
\newcommand{\Vast}{\bBigg@{5}}
\makeatother

\title{A collection of definitions and fundamentals for a design-oriented inductor model}

\author{\IEEEauthorblockN{1\textsuperscript{st} Andr\'es Vazquez Sieber}
\IEEEauthorblockA{\textit{* Departamento de Electr\'onica} \\
\textit{Facultad de Ciencias Exactas, Ingenier\'ia y Agrimensura} \\
\textit{Universidad Nacional de Rosario (UNR)} \\
\textit{** Grupo Simulaci\'on y Control de Sistemas F\'isicos} \\
\textit{CIFASIS-CONICET-UNR} \\
Rosario, Argentina \\
avazquez@fceia.unr.edu.ar}
\and
\IEEEauthorblockN{2\textsuperscript{nd} M\'onica Romero}
\IEEEauthorblockA{\textit{* Departamento de Electr\'onica} \\
\textit{Facultad de Ciencias Exactas, Ingenier\'ia y Agrimensura} \\
\textit{Universidad Nacional de Rosario (UNR)} \\
\textit{** Grupo Simulaci\'on y Control de Sistemas F\'isicos} \\
\textit{CIFASIS-CONICET-UNR} \\
Rosario, Argentina \\
mromero@fceia.unr.edu.ar}
}

\maketitle

\begin{abstract}
This paper defines and develops useful concepts related to the several kinds of inductances employed in any comprehensive design-oriented ferrite-based inductor model, which is required to properly design and control high-frequency operated electronic power converters. It is also shown how to extract the necessary parameters from a ferrite material datasheet in order to get inductor models useful for a wide range of core temperatures and magnetic induction levels. 
\end{abstract}


\section{Introduction}

\IEEEPARstart{F}{e}rrite-core based low-frequency-current biased inductors are commonly found, for example, in the LC output filter of voltage source inverters (VSI) or step-down DC/DC converters. Those inductors have to effectively filter a relatively low-amplitude high-frequency current being superimposed on a relatively large-amplitude low-frequency current. It is of paramount importance to design these inductors in a way that a minimum inductance value is always ensured which allows the accurate control and the safe operation of the electronic power converter.
In order to efficiently design that specific type of inductor, a method to find the required minimum number of turns $N_{min}$ and the optimum air gap length $g_{opt}$ to obtain a specified inductance at a certain current level is needed. This method has to be based upon an accurate inductor model, for which certain inductances and properties need to be defined and explained. Also, these inductance definitions needs to be parametrized, among other things, according to the specific ferrite material employed in the core. 

The problem of designing such kind of inductors has been widely treated in literature \cite{TransformerInductorDesign:McLyman}, \cite{PowerElectronics:Mohan}, \cite{High-FrequencyMagneticComponents:Kazimierczuk}, \cite{InductorsTransformersPowerElectronics:Bossche}. At the same time, there are many well established definitions of core permeability and inductance \cite{InductorsTransformersPowerElectronics:Bossche}, \cite{SoftFerrites:Snelling}, \cite{MagneticMaterialsApplications:Heck} according to the actual inductor operating condition. However, it seems that this variety of inductance definitions can be better exploited in order to enhance the inductor design process. In this paper, some specific inductance definitions are revisited and presented under a suitable context for the power electronic practitioner.
    %efinitions,from the authors' perspective, it has not been well addressed how the several types of defined inductances [][][] impact and shape the computation of the actual inductance value when the inductor is under different operating condition. Although there are all kind of magenetic permeability and inductance definitions, which are well stablished in literature [][][], in this paper some of them are revisited and presented in a more suitable way for the power electronic practisioner.
A design-oriented inductor model can be based on the core magnetic model described in this paper which allows to employ the concepts of reversible inductance $L_{\hat{rev}}$, amplitude inductance $L_a$ and initial inductance $L_i$, to further develop an optimized inductor design method. Those inductance definitions rely on their respective core permeabilities, which in this paper are also revisited, contextualized and obtained for two specific ferrite materials: TDK-EPCOS N27 and TDK-EPCOS N87. 

This paper is organized as follows. In section \ref{sec:MagneticCircuitModel}, a general magnetic core model is described and its associated permeabilities are introduced. In section \ref{sec:InductanceDefinitions}, definitions of several types of inductances are presented. Section \ref{sec:PermeabilityModels} shows how to obtain the previously defined permeabilities from the ferrite material datasheet. Section \ref{sec:ConsiderationsReversibleInductance} presents some useful properties of the reversible inductance that could be needed to justify the selection criterion of the inductance value as well as $N_{min}$ and $g_{opt}$. Finally, conclusions are presented in section \ref{sec:Conclusions}.

\section{Magnetic circuit model} \label{sec:MagneticCircuitModel}

\begin{figure}
					\centering
		\includegraphics[width=8.5cm]{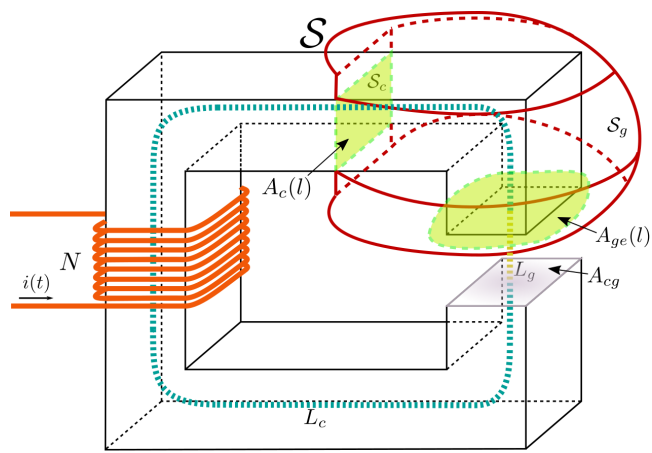}
       \caption{General magnetic circuit}
       \label{fig:Magnetic_circuit}
\end{figure}

In this section, we obtain a model for the general magnetic circuit considered in Figure \ref{fig:Magnetic_circuit} using an approach depending on integration along the mean magnetic path into the ferrite core, $L_c$ and into the air gap, $L_g$ \cite{SoftFerrites:Snelling}. 
Suppose that current $i(t)$ can be easily decomposed into a) a component denoted as $i_{LF}(t)$ at a relatively low frequency $f_{LF}$ and b) a component denoted as $i_{HF}(t)$ at a relatively high frequency $f_{HF}$, with $f_{LF} \ll f_{HF}$. At a certain time $\hat t_{LF}$, $i_{LF}$ reaches its peak value $\hat i_{LF}$ and then  we have
\begin{align}
			\label{eq:t_interval}	
		i(t) \approx \hat i_{LF}+i_{HF}(t) \quad			
			t \in \left[ \hat t_{LF} -\frac{1}{f_{HF}}, \hat t_{LF} +\frac{1}{f_{HF}} \right]
	\end{align}
	In such a situation, Ampere's law relates the frequency components of current $i(t)$ with their corresponding magnetic field strength $H$ components as follows
			\begin{align*}
		\oint \vec{H}(t,l) \vec{dl}&=\int_{L_c \cup L_g} \left[ \hat H_{LF}(l)+H_{HF}(t,l) \right] dl \\
		&= N \left[ \hat i_{LF}+i_{HF}(t) \right] 
	\end{align*}						
	where $\vec{dl}$ is the path vector, parallel to $\vec{H}$. Separating the frequency components yields
			\begin{align}
			\label{eq:int_HLF}
		\int_{L_c} \hat H_{LF}(l) dl + \int_{L_g} \hat H_{LF}(l) dl& = N \hat i_{LF} \\
		\int_{L_c} H_{HF}(t,l) dl + \int_{L_g} H_{HF}(t,l) dl &= N i_{HF}(t)	\notag \\	
	\label{eq:int_HHF}
			\int_{L_c} \Delta H_{HF}(l) dl + \int_{L_g} \Delta H_{HF}(l) dl &= N \Delta i_{HF}		
	\end{align}
	where $\Delta H_{HF}$ is the amplitude of the field strength excursion due to $\Delta i_{HF}$, the amplitude of the high-frequency current excursion during $\frac{1}{f_{HF}}$.	
			
\begin{figure}
					\centering
		\includegraphics[width=8.5cm]{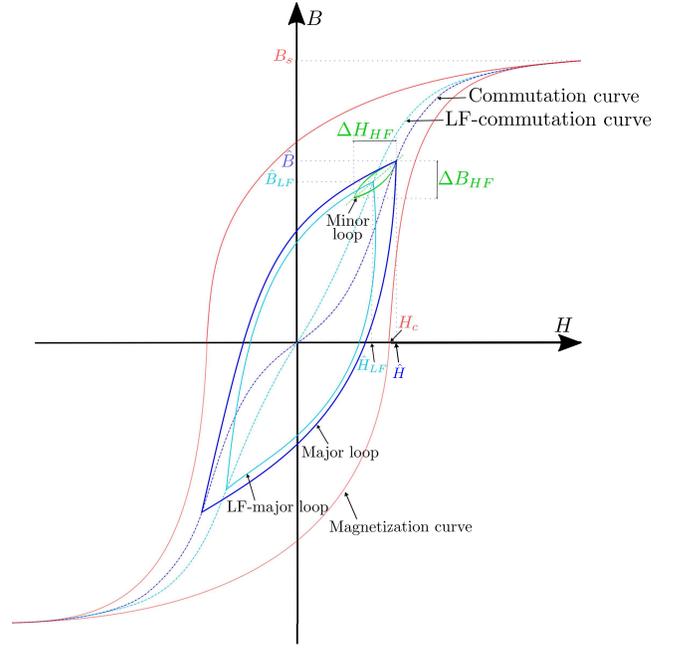} 
       \caption{Ferrite magnetization curve}
       \label{fig:Magnetic_loop}
\end{figure}

	The magnetic induction $B(t,l)=\hat B_{LF}(l)+B_{HF}(t,l)$ and its peak-to-peak variation $\Delta B_{HF}$ determine the peak induction $\hat B(l)=\hat B_{LF}(l)+\frac{\Delta B_{HF}(l)}{2}$. These are related to their corresponding field strength $\hat H_{LF}(l)$, $H_{HF}(t,l)$, $\Delta H_{HF}(l)$ and $\hat H(l)$ according to the medium permeability. Having the air gap paramagnetic properties, along $L_g$ simply hold
	\begin{align}
	\label{eq:muo}
			\frac{\hat B_{LF}(l)}{\hat H_{LF}(l)}= \frac{B_{HF}(t,l)}{H_{HF}(t,l)}= \frac{\Delta B_{HF}(l)}{\Delta H_{HF}(l)}=\frac{\hat B(l)}{\hat H(l)}=\mu_0			
	\end{align}			
where $\mu_0$ is the vacuum permeability. In the magnetic core path $L_c$, those relationships depend on the shape of the ferrite magnetization curve which is shown in Figure \ref{fig:Magnetic_loop}. It is also the specific major loop that characterizes the behaviour of the ferrite when its magnetic induction evolution spans the two extreme points $\pm B_s$, being $B_s$ the saturation induction. In any inductor, although this situation can be reached with a current $i_{LF}$ having a sufficiently high $\hat i_{LF}$, the actual $\hat i_{LF}$ has to be set well below that value since beyond that induction level the core ferrimagnetic properties become severely affected. Starting from a demagnetized core, as $\hat i_{LF}$ is gradually increased from zero towards the maximum value causing saturation, the tipping points ($\hat B_{LF}$,$\hat H_{LF}$ and $-\hat B_{LF}$,$-\hat H_{LF}$) of the ever increasing LF-major loops describe a LF-commutation curve which is also partly shaped by the current magnitude of $\Delta B_{HF}$, due to the memory properties of the ferrite material.  

For any of the points pertaining to that LF-commutation curve, the LF-amplitude permeability $\mu_{a}^{LF}$ is defined as
	\begin{align*}
			\mu_{a}^{LF} \left( \hat B_{LF}, \Delta B_{HF} \right) =\left. \frac{1}{\mu_0}\frac{\hat B_{LF}}{\hat H_{LF}} \right|_{\Delta B_{HF}}
	\end{align*}
because it relates only the low frequency amplitude of the magnetic induction and field strength in the ferrite material when the low frequency $i_{LF}$ takes also its amplitude value $\hat i_{LF}$. 
The magnetic induction generated by $i_{LF}$ will not vary much from $\hat B_{LF}$ while $t$ is into the time span defined in \eqref {eq:t_interval}. Hence during $\frac{1}{f_{HF}}$, $i_{HF}(t)$ will produce an approximately closed minor magnetic loop of amplitude ($\Delta B_{HF}, \Delta H_{HF}$) starting and ending in the neighbourhood of $\hat B_{LF}$, as it is shown in Figure \ref{fig:Magnetic_loop}. The incremental permeability $\mu_{\hat \Delta}$ at that quasi-static induction level $\hat B_{LF}$ is then defined as
   \begin{align*}
			\mu_{\hat \Delta} \left( \hat B_{LF}, \Delta B_{HF} \right)=\left. \frac{1}{\mu_0}\frac{\Delta B_{HF}}{\Delta H_{HF}} \right|_{\hat B_{LF}}
	\end{align*}
Consequently, on $L_c$ holds
 \begin{align}
\label{eq:HLF_HHF}
		\hat H_{LF}(l) &= \frac{\hat B_{LF}(l)}{\mu_0 \mu_a^{LF} \left( \hat B_{LF}, \Delta B_{HF} \right) } \\
		\Delta H_{HF}(l) &= \frac{\Delta B_{HF}(l)}{\mu_0 \mu_{\hat \Delta} \left(\hat B_{LF},\Delta B_{HF} \right)} 	
	\end{align}
	
	If $\Delta B_{HF}$ is made sufficiently small, then $\mu_{\hat \Delta}$ and the LF-commutation curve start to be practically independent of $\Delta B_{HF}$. At this point, on the one hand the existing linear relationship between $B_{HF}$ and $H_{HF}$ is captured by the so-called reversible permeability at $\hat B_{LF}$, $\mu_{\hat {rev}}$ 
\begin{align*}
		\mu_{\hat {rev}} \left( \hat B_{LF} \right) =\lim_{\substack{\Delta B_{HF}  \to 0}} \mu_{\hat \Delta} \left( \hat B_{LF}, \Delta B_{HF} \right)
\end{align*}
On the other hand, the LF-commutation curve tends to the regular commutation curve and their respective amplitude permeabilities are related as
\begin{align*}
		\mu_{a} \left( \hat B \right)&=\frac{1}{\mu_0}\frac{\hat B}{\hat H}=\lim_{\substack{\hat B_{LF} \to \hat B \\ \Delta B_{HF} \to 0}} \mu_{a}^{LF} \left( \hat B_{LF}, \Delta B_{HF} \right) \\
		&= \mu_{a} \left( \hat B_{LF} \right)		
\end{align*}
Now making $\hat B_{LF} \to 0$ due to $\hat i_{LF} \to 0$, the initial permeability $\mu_i$ is defined as 
 \begin{align*}
		\mu_i =\lim_{\substack{\hat B_{LF}  \to 0}} \mu_a \left( \hat B_{LF} \right)=\mu_{\hat{rev}} \left( \hat B_{LF}=0 \right)
\end{align*}
	Note that in the core, the relationship between $B$ and $H$ depends not only on the ferrite magnetic characteristics but also on the way in which $B$ evolves with time.  
	
	In the magnetic circuit of Figure \ref{fig:Magnetic_circuit}, a closed surface $\mathcal{S}=\mathcal{S}_c \cup \mathcal{S}_g$ that intersects both $L_c$ and $L_g$ paths will satisfy according to Gauss' law that
	\begin{align}
\label{eq:Fluxes}		
		\iint_{\mathcal{S}_c} \vec{B}(t,l) \vec{dS} = \iint_{\mathcal{S}_g} \vec{B}(t,l) \vec{dS} 
		\end{align}
	where $\mathcal{S}_c$ is a core cross-section perpendicular to $L_c$, $\mathcal{S}_g$ is the remaining surface of $\mathcal{S}$ crossing the air gap and $\vec{dS}$ is the area vector of $\mathcal{S}$. The left side of \eqref{eq:Fluxes} is the core magnetic flux $\Phi_c$ since the magnetic induction there is mainly concentrated into $\mathcal{S}_c$ because $\mu_a, \mu_{\hat \Delta} \gg 1$. All the magnetic induction in the air gap will then pass through $\mathcal{S}_g$, thus the right side of \eqref{eq:Fluxes} is the air gap magnetic flux $\Phi_g$. Consequently, $\Phi_c=\Phi_g=\hat \Phi_{LF}+\Phi_{HF}(t)$. $L_c$ passes perpendicular through the center of $\mathcal{S}_c$, so the magnetic induction along $L_c$ will be approximately an average of that existing inside $\mathcal{S}_c$ and equal to
	\begin{align}
	\label{eq:Phi_c}
		\hat B_{LF}(l) +  B_{HF}(t,l)=\frac{\hat \Phi_{LF}+\Phi_{HF}(t)}{A_c(l)}
	\end{align}
	where $A_c(l)$ is the area of $\mathcal{S}_c$ at a certain point $l \in L_c$.
	Around the air gap, the magnetic induction is far more nonuniform in $\mathcal{S}_g$ than in $\mathcal{S}_c$ due to the fringing flux. Thus, the mean induction on $L_g$ can be quite different from the actual values at the edges of the gap, but being a paramagnetic region, it suffices to propose an effective gap area $A_{ge}(l)$ with $l \in L_g$, as if all the induction were there concentrated.    
		\begin{align}
		\label{eq:Phi_g}
		\hat B_{LF}(l) + B_{HF}(t,l)=\frac{\hat \Phi_{LF}+\Phi_{HF}(t)}{A_{ge}(l)}
	\end{align}
Note that $A_{ge}(l)$ is approximately equal to $A_{cg}$, the core cross-section in contact with the air gap, if its length is much smaller than the linear dimensions characterizing $A_{cg}$. Given that the winding turns $N$ embrace practically all $\Phi_c$, it follows that the linkage flux $\Psi$ is 
	\begin{align}
	\label{eq:Psi}
		\hat \Phi_{LF}+ \Phi_{HF}(t) =\frac{\hat \Psi_{LF}+ \Psi_{HF}(t)}{N}			
	\end{align}
The peak linkage flux $\hat \Psi$ and magnetic induction $\hat B(l)$ in the ferrite core are
	\begin{align}
	\label{eq:peak_Psi}
		\hat \Psi &=\hat \Psi_{LF}+ \frac{\Delta \Psi_{HF}}{2} \\
		\hat B(l) &=\frac{\hat \Psi}{NA_c(l)}	\notag 		
	\end{align}		

\section{Inductance definitions} \label{sec:InductanceDefinitions}

Combining \eqref{eq:int_HLF}, \eqref{eq:int_HHF}, \eqref{eq:muo}, \eqref{eq:HLF_HHF}, \eqref{eq:Phi_c}, \eqref{eq:Phi_g} and \eqref{eq:Psi} yield the LF-amplitude inductance, $L_a^{LF}$ and the incremental inductance, $L_{\hat \Delta}$
	\begin{align}
	\label{eq:LaLF_def}		 
		 L_a^{LF} \left( \hat \Psi_{LF}, \Delta \Psi_{HF} \right) &=\left. \frac{ \hat \Psi_{LF}}{ \hat i_{LF}} \right|_{ \Delta \Psi_{HF}} \notag \\
		&= \frac{N^2}{\mathcal{R}_{c_a}^{LF} \left( \hat \Psi_{LF}, \Delta \Psi_{HF} \right) +\mathcal{R}_g}  \\
		\label{eq:Ldelta}	
	L_{\hat \Delta} \left( \hat \Psi_{LF}, \Delta \Psi_{HF} \right) &= \left. \frac{ \Delta \Psi_{HF}}{ \Delta i_{HF}} \right|_{ \hat \Psi_{LF}} \notag \\
	&= \frac{N^2}{\mathcal{R}_{c_{\hat \Delta}}\left( \hat \Psi_{LF}, \Delta \Psi_{HF} \right)+\mathcal{R}_g}
	\end{align}
	with
	\begin{align}
	\label{eq:RcaLF}
	\mathcal{R}_{c_a}^{LF} \left( \hat \Psi_{LF}, \Delta \Psi_{HF} \right) &=\int_{L_c} \frac{dl}{\mu_0\mu_a^{LF} \left( \frac{\hat \Psi_{LF}}{NA_c}, \frac{\Delta \Psi_{HF}}{NA_c} \right) A_c(l)} \\
	\label{eq:Rcdelta}
	\mathcal{R}_{c_{\hat \Delta}} \left( \hat \Psi_{LF}, \Delta \Psi_{HF} \right) &= \int_{L_c} \frac{dl}{\mu_0\mu_{\hat \Delta} \left( \frac{\hat \Psi_{LF}}{NA_c}, \frac{\Delta \Psi_{HF}}{NA_c} \right) A_c(l)} \\
	\label{eq:Rg_int}
	\mathcal{R}_g &= \int_{L_g} \frac{dl}{\mu_0 A_{ge}(l)}
	\end{align}	
	$\mathcal{R}_{c_a}^{LF}$, $\mathcal{R}_{c_{\hat \Delta}}$ and $\mathcal{R}_g$ are the core LF-amplitude reluctance, core incremental reluctance and the air gap reluctance respectively.
				
Considering the situation where $\Delta \Psi_{HF} \to 0$, $\mu_a$ and $\mu_{\hat {rev}}$ define the amplitude and reversible inductances at $\hat \Psi_{LF}$, $L_a$ and $L_{\hat{rev}}$, as
\begin{align}
\label{eq:La_def}
		  L_a \left( \hat \Psi_{LF} \right) &=\lim_{\substack{\hat \Psi_{LF}  \to \hat \Psi \\ \Delta \Psi_{HF}  \to 0}} L_a^{LF} \left( \hat \Psi_{LF}, \Delta \Psi_{HF} \right) \notag \\
		&=\frac{N^2}{\mathcal{R}_{c_a} \left( \hat \Psi_{LF} \right) +\mathcal{R}_g}  \\
		L_{\hat {rev}} \left( \hat \Psi_{LF} \right) &=\lim_{\substack{\Delta \Psi_{HF}  \to 0}} L_{\hat \Delta} \left( \hat \Psi_{LF}, \Delta \Psi_{HF} \right) \notag \\
		&=\frac{N^2}{\mathcal{R}_{c_{\hat{rev}}} \left( \hat \Psi_{LF} \right) +\mathcal{R}_g} \notag  \\
		\label{eq:Rca}
	\mathcal{R}_{c_a} \left( \hat \Psi_{LF} \right) &=\int_{L_c} \frac{dl}{\mu_0\mu_a \left( \frac{\hat \Psi_{LF}}{NA_c} \right) A_c(l)} \\
		\label{eq:Rc_rev}
\mathcal{R}_{c_{\hat {rev}}} \left( \hat \Psi_{LF} \right) &= \int_{L_c} \frac{dl}{\mu_0\mu_{\hat {rev}} \left( \frac{\hat \Psi_{LF}}{NA_c} \right) A_c(l)}
\end{align}
$\mathcal{R}_{c_a}$ and $\mathcal{R}_{c_{\hat{rev}}}$ are the core amplitude reluctance and the core reversible reluctance respectively.

$L_a$ and $L_{\hat{rev}}$ usually have dissimilar values at a same $\hat \Psi_{LF}$ and vary differently as $\hat \Psi_{LF}$ increases from zero to relatively high values. It is then important to find a common situation to relate and relativize their current values with. In a demagnetized material, $\mu_a$ and $\mu_{\hat{rev}}$ coincide at the origin which means that $L_a$ and $L_{\hat{rev}}$ converge to the initial inductance $L_i$  
\begin{align}
\label{eq:La_Lrev_conv_Li}
		L_i & = \lim_{\substack{\hat \Psi_{LF}  \to 0}} L_a \left( \hat \Psi_{LF} \right)=L_{\hat {rev}} \left( \hat \Psi_{LF}=0 \right)=\frac{N^2}{\mathcal{R}_{c_i}+\mathcal{R}_g} \\
		\label{eq:Rci}
	\mathcal{R}_{c_i} &=\int_{L_c} \frac{dl}{\mu_0\mu_i  A_c(l)}
\end{align}		
being $\mathcal{R}_{c_i}$ the core initial reluctance.
Note that only $\mu_i$ does not vary with the core cross-sectional area $A_c(l)$ along the magnetic path. However, $\mu_i$ as well as $\mu_a$ and $\mu_{\hat {rev}}$ do depend heavily on the core temperature, as is modeled in the next section.
  
\section{Permeability models} \label{sec:PermeabilityModels}

The dependence of $\mathcal{R}_{c_a}$ in \eqref{eq:Rca} and $\mathcal{R}_{c_i}$ in \eqref{eq:Rci} from core temperature $T_c$ and magnetic induction $\hat B_{LF}=\hat B$, in each specific part of the core, is addressed when the corresponding functions $\mu_a\left( \hat B_{LF}, T_{c}\right)$ and $\mu_i \left( T_{c} \right)$ 
are extracted from the ferrite material datasheet \cite{FerriteN27:TDK-EPCOS} \cite{FerriteN87:TDK-EPCOS}. Permeability $\mu_a$ is a function of magnetic induction amplitude $\hat B$ and core temperature (3-D lookup table) and permeability $\mu_i$ is a function of core temperature (2-D lookup table). 
To get the best accuracy in the inductor model%\cite{MyPaper:AVS_MR}
, the $\mu _a$ curve given by the ferrite manufacturer should have been obtained at a frequency close to $f_{LF}$.

The temperature and induction dependence of $\mathcal{R}_{c_{\hat{rev}}}$ in \eqref{eq:Rc_rev} is subjected to find $\mu_{\hat{rev}} \left( \hat B_{LF}, T_{c}\right)$. In \cite{ComputationMinorLoops:Esguerra} it is concluded that the commutation curve coincides with the so-called initial magnetization curve for soft ferrite materials, that is $(B_{DC},H_{DC})=(\hat B,\hat H)$. This means that $\mu_{\hat{rev}}$ is equal to DC-biased $\mu_{rev}$ which can be extracted from a graph or as a function of DC-bias field strength $H_{DC}$. 
That curve may not be given in datasheets for a particular ferrite material or for the core temperatures at which $\mu_{\hat{rev}}$ has to be obtained, but even if it were available it should be put in terms of $\hat B_{LF}$ to be employed in \eqref{eq:Rc_rev}. To overcome these limitations, we use a permeability model directly relating DC-biased $\mu_{rev}$ with DC-bias magnetic induction $B_{DC}$ \cite{ModellingHysteresisLoops:Esguerra}, where all its parameters at the desired core temperature can be entirely obtained from any ferrite datasheet, in the way it is next explained. This approach has been experimentally validated for many ferrite materials operating at different temperatures \cite{MajorHysteresisLoopsPermeability:Esguerra} and it is currently employed by major ferrite manufacturers \cite{MagneticsDesignTool:Esguerra}.

Let us first consider the empirical models that curvefit the upper (u) and lower (l) branches of the dynamic magnetization (B-H) curve of Figure \ref{fig:Magnetic_loop} \cite{ModellingHysteresisLoops:Esguerra}, 
	\begin{align}
		\label{eq:BH_regression_model_upper}
		H_u(B)&=\frac{B}{\mu_0 \mu_{c}}\frac{1}{1-\left(\frac{B}{B_{s}}\right)^{a_u}}-H_{c}	\\
	\label{eq:BH_regression_model_lower}
		H_l(B)&=\frac{B}{\mu_0 \mu_{c}}\frac{1}{1-\left(\frac{B}{B_{s}}\right)^{a_l}}+H_{c}
	\end{align}	 
 with the positive parameters: coercive field strength $H_c$, coercive permeability $\mu_{c}$ and squareness coefficients $a_u$ and $a_l$ for each branch. Supposing that $a_l \approx a_u$ and being $\hat B=\hat B_{LF}$, $\mu_{\hat{rev}} \left( \hat B_{LF}, T_{c}\right)$ can be expressed as \cite{ModellingHysteresisLoops:Esguerra}
	\begin{align}
	\label{eq:invMuRev}
		\mu_{\hat{rev}} \left( \hat B_{LF}, T_{c} \right)&=\vast\{ \frac{1+(a_l-1) \left(\frac{\hat B_{LF}}{B_s}\right)^{a_l}}{\left[1-\left(\frac{\hat B_{LF}}{B_s}\right)^{a_l}\right]^2}\frac{1}{\mu_c} \notag \\
		&+\frac{b_o}{\left(1-\frac{\hat B_{LF}}{B_s}\right)\left[2-\left(1-\frac{\hat B_{LF}}{B_s}\right)^{a_o}\right]} \vast\}^{-1} \\
			a_o&=\frac{b_o B_s}{\mu_0 H_c} 	\quad \quad  b_o=\frac{1}{\mu_i}-\frac{1}{\mu_c}		\notag
	\end{align}
Apart from $\mu_i$, $\mu_{\hat{rev}}$ depends on $T_{c}$ through $B_s$, $H_c$, $a_l$ and $\mu_c$ and thus \eqref{eq:BH_regression_model_lower} has to be numerically fitted for each particular $T_{c}$. The fitting data is obtained from the 3-D lookup table $H(B,T=T_{c})$ based on the corresponding curves from the ferrite material datasheet \cite{FerriteN27:TDK-EPCOS} \cite{FerriteN87:TDK-EPCOS}. 

The starting guess points for the fitting process are extracted from 2-D lookup tables $B_s^*(T)$, $H_c^*(T)$, $a_l^*(T)$ and $\mu_c^*(T)$%, with $T=T_{cz}$
. $B_s^*(T)$ and $H_c^*(T)$ are built to linearly interpolate the two saturation induction and coercive field strength values ($B_{s1}$, $B_{s2}$, $H_{c1}$ and $H_{c2}$ respectively), that are stated at the two corresponding temperatures $T_1$, $T_2$, in the datasheet of the magnetic material. Lookup tables $a_l^*(T)$ and $\mu_c^*(T)$ are conformed in the following way. Let $H_{11}(B_{11},T_1)$ and $H_{12}(B_{12},T_1)$ be the field strength at two different induction levels from the lower branch of the B-H curve at temperature $T_1$ given by the datasheet. The value $B_{11}$ could be from the "linear" region of the curve, while $B_{12}$ could be taken from the "knee" between "linear" and "saturation" regions of the curve at temperature $T_1$. The estimations of coefficients $a_l$ and $\mu_{c}$ from \eqref{eq:BH_regression_model_lower} at temperature $T_1$, $a_{l1}^*$ and $\mu_{c1}^*$ respectively, are found numerically solving
		\begin{align*}
		\frac{1-\left( \frac{B_{11}}{B_{s1}} \right)^{a_{l1}^*}}{1-\left( \frac{B_{12}}{B_{s1}} \right)^{a_{l1}^*}}&=\frac{H_{12}-H_{c1}}{H_{11}-H_{c1}}\frac{B_{11}}{B_{12}}	\\
		\mu_{c1}^*&=\frac{1}{\mu_0}\frac{B_{11}}{H_{11}-H_{c1}}\frac{1}{1-\left(\frac{B_{11}}{B_{s1}}\right)^{a_{l1}^*}}
	\end{align*}
	
\begin{figure}
       \centering
		\includegraphics[width=8.5cm]{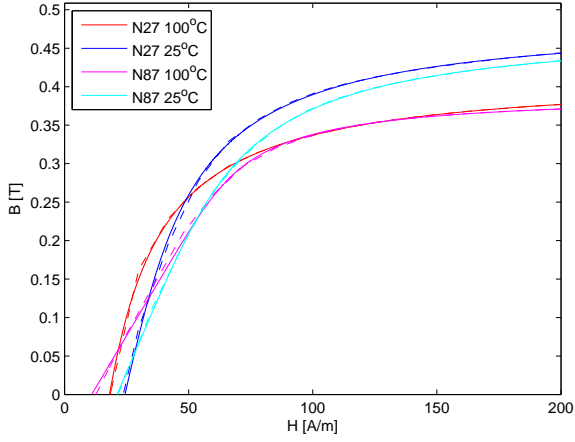}
       \caption{Curve-fitting of the lower branch of the magnetization loops (first quadrant) for materials N27 and N87. Dashed lines are interpolated data from datasheet}
       \label{fig:Loops_N27_N87}
\end{figure}

		Using the B-H curve at temperature $T_2$, $a_{l2}^*$ and $\mu_{c2}^*$ can be also obtained following a similar reasoning. Finally, $a_l^*(T)$ and $\mu_c^*(T)$ are built to linearly interpolate $a_{l1}^*,a_{l2}^*$ and $\mu_{c1}^*,\mu_{c2}^*$ respectively. Figure \ref{fig:Loops_N27_N87} shows the fitting goodness of \eqref{eq:BH_regression_model_lower} for TDK-EPCOS ferrite materials N27 and N87 while their corresponding parameters for using \eqref{eq:invMuRev} are summarized in Table \ref{table:param_urev}. To get the best accuracy in the inductor model, the magnetization curve given by the ferrite manufacturer should have been obtained at a frequency close to $f_{HF}$.	
\begin{table}[ht]
\caption{Reversible permeability $\mu_{\hat{rev}}$ model parameters (Equation \eqref{eq:invMuRev})} 
\centering 
\begin{tabular}{c c c c c c c} 
\hline\hline 
Material & $T_c$ [$^oC$] & $a_l$ & $H_c$ [$A/m$] & $\mu_c$ & $\mu_i$ & $B_s$ [$T$]\\ [0.5ex] 
\hline 
N27 & 100 & 1.25 & 18.12  & 14079 & 3231 & 0.4165\\ 
N27 & 25 & 2.00 & 24.35  & 11154 & 1700 & 0.4895\\
N87 & 100 & 8.00 & 10.94  & 4330 & 3976 & 0.3925\\
N87 & 25 & 3.78 & 21.17  & 6014 & 2210 & 0.4803\\ [1ex] 
\hline 
\end{tabular}
\label{table:param_urev} 
\end{table}
\section{Considerations on the reversible inductance} \label{sec:ConsiderationsReversibleInductance}

Recall that if $\Delta \Psi_{HF} \to 0$ we can consider
\begin{align*}
L_{\hat{rev}} \left( \hat \Psi \right) =L_{\hat{\Delta}} \left( \hat \Psi_{LF},\Delta \Psi_{HF} \right) = L_{\hat{rev}} \left( \hat \Psi_{LF} \right)
\end{align*}
It is important to remark that in this situation $L_{\hat{rev}}$ is the minimum value of reversible inductance arising along the whole symmetric major loop having $\pm \left (\hat  H_{LF}, \hat  B_{LF} \right ) $ as tipping points. This is in fact proved by taking into consideration  
the upper branch of the particular $\pm \left (\hat  H_{LF}, \hat  B_{LF} \right ) $ major loop which can be described in terms of \eqref{eq:BH_regression_model_upper}-\eqref{eq:BH_regression_model_lower} as \cite{ComputationMinorLoops:Esguerra}
\begin{align}    
\label{eq:Blf_Hlf_upper branch}
		H^{LF}_u(B)&=H_u(B) +  \frac{\left [\hat H_{LF} - H_u(\hat B_{LF}) \right ]^{\beta}}{\left [ H_l(\hat B_{LF}) - \hat H_{LF} \right ]^{\alpha}} 	\\
				\alpha &=\frac{B-\hat B_{LF}}{2\hat B_{LF}} \quad \beta = \frac{B+\hat B_{LF}}{2\hat B_{LF}} \notag
	\end{align}	
and its first derivative, the inverse of the so-called differential permeability $\mu _d$ \cite{InductorsTransformersPowerElectronics:Bossche}
\begin{align*}
\frac{d H^{LF}_u}{d B}=\frac{1}{\mu _d}
\end{align*}
Since $\mu_{rev} < \mu_d$ \cite{InductorsTransformersPowerElectronics:Bossche} it is sufficient to show that $\frac{d \mu_d}{d B}<0$ to conclude that $L_{rev}$ is decreasing when $B$ takes values from zero to $\hat B_{LF}$ and hence $L_{\hat{rev}}$ is the absolute minimum. Accordingly,
\begin{align}
\frac {d \mu _d}{d B}&= - {\mu_d}^2 \frac{d^2 H^{LF}_u}{d B^2}		\notag \\
\label{eq:d2H/dB2}
\frac{d^2 H^{LF}_u}{d B^2} &= \frac{a_u \left( \frac{B}{B_s} \right)^{a_u} \left[ 1 - \left( \frac{B}{B_s} \right)^{a_u} + a_u + a_u \left( \frac{B}{B_s} \right)^{a_u} \right] }{\mu_0 \mu_c \left[ 1- \left( \frac{B}{B_s} \right)^{a_u} \right]^3 B}  	\notag \\
&+ \left[ \frac{\ln \frac{H_l(\Hat B_{LF})-\hat H_{LF}}{\hat H_{LF} - H_u(\Hat B_{LF})}  }{2 \hat B_{LF}} \right]^2 \frac{\left[ \hat H_{LF} - H_u(\hat B_{LF})\right]^{\beta}}{\left[H_l(\hat B_{LF}) - \hat H_{LF} \right]^{\alpha}}
\end{align}
$\hat H_{LF}$ is inside the area delimited by the largest major loop, the magnetization curve, that is described by \eqref{eq:BH_regression_model_upper}-\eqref{eq:BH_regression_model_lower}. Consequently, \eqref{eq:d2H/dB2} is positive for $B \in [ 0,\hat B_{LF}] $ and thus $\mu_d$ is ever decreasing for increasing values of $B>0$.   

Now suppose that gradually $\Delta \Psi_{HF}$ is increased and $\hat \Psi_{LF}$ is decreased in such a way that $\hat \Psi$ remains unchanged, implying that $\hat B$ and $\hat H$ are unmodified in all parts of the core. This scenario brings into existence increasingly asymmetric minor loops in the $B-H$ plane with tipping points 
\begin{align*}
\hat B &= \hat B_{LF}+\frac{\Delta B_{HF}}{2} \\
\hat H &=\hat H_{LF} +k_H \Delta H_{HF} \\
\hat B -\Delta B_{HF} &=\hat B_{LF}-\frac{\Delta B_{HF}}{2} \\
\hat H - \Delta H_{HF} &= \hat H_{LF} - (1-k_H) \Delta H_{HF} 
\end{align*}
being $k_H \in [0.5, 1)$ the magnetic field symmetry factor. 
Considering that along a general magnetic loop $\mu_d$ increases as $B$ decreases, it can be stated that
\begin{align*}
\left. \hat B - \frac{d B}{d H} \right|_{\hat B} \Delta H_{HF} \geq \hat B - \Delta B_{HF} %\\
\end{align*}
and hence   
\begin{align*}
\mu_{\hat{rev}} \left( \hat B \right) \leq \mu_d \left( \hat B \right) \leq \mu_{\hat \Delta} \left( \hat B_{LF},\Delta B_{HF} \right) 
\end{align*}
Inside the minor loop, we can define the minor-loop amplitude permeability $\mu_a^{MN}$ as
\begin{align*}
\mu_a^{MN}=\frac{1}{\mu_0} \frac{\hat B - \hat B_{LF}}{\hat H - \hat H_{LF}}=\frac{1}{\mu_0} \frac{\Delta B_{HF}}{2 k_H \Delta H_{HF}}=\frac{1}{2 k_H} \mu_{\hat \Delta}
\end{align*}
and to note that it holds $\mu_{\hat{rev}} \left( \hat B \right) < \mu_{\hat{rev}} \left( \hat B_{LF} \right)$, since when $B$ is far from the origin, $\mu_{\hat{rev}}$ decreases as $B$ increases. Consequently, 
\begin{align*}
\hat B_{LF} + \mu_0 \mu_{\hat{rev}} \left( \hat B_{LF} \right) k_H \Delta H_{HF} \geq \hat B_{LF} + \frac{\Delta B_{HF}}{2}
\end{align*}
and hence
\begin{align*}
\mu_{\hat{rev}} \left( \hat B_{LF} \right) \geq \mu_a^{MN} \leq \mu_{\hat \Delta}
\end{align*} 
If the minor loop keeps some degree of symmetry around $\left( \hat B_{LF}, \hat H_{LF} \right)$, i.e. $k_H \approx 0.5$, then $\mu_a^{MN} \approx \mu_{\hat \Delta}$ and hence 
\begin{align*}
\mu_{\hat{rev}} \left( \hat B \right) \leq \mu_{\hat \Delta} \left( \hat B_{LF},\Delta B_{HF} \right) \leq \mu_{\hat{rev}} \left( \hat B_{LF} \right)
\end{align*}
However, that minor loop with tipping points
\begin{align*}
 \left( \hat B,\hat H \right); \left( \hat B-\Delta B_{HF},\hat H - \Delta H_{HF} \right)
\end{align*}
 and an enclosing symmetric major loop with tipping points
\begin{align*}
 \left( \hat B,\hat H \right); \left( -\hat B,-\hat H \right)
\end{align*}
 coincide in the vicinity of their uppermost tipping point $\left( \hat B,\hat H \right)$ \cite{HighOrderHystereticLoops:Harrison} \cite{ModellingHysteresisLoops:Esguerra}. 
Hence, $\mu_{\hat{rev}}(\hat B)$ at the existing minor loop is equal to $\mu_{\hat{rev}}(\hat B)$ at that hypothetical major loop. In fact, $\mu_{rev} \left( B_{DC} \right)$ \cite{ModellingHysteresisLoops:Esguerra}, from which \eqref{eq:invMuRev} is particularly derived, depends on the current magnetic induction value regardless its previous evolution \cite{MagneticMaterialsApplications:Heck}. 
Note that from the minor loop standpoint, $\hat \Psi$ is given by \eqref{eq:peak_Psi}, but it can be also put in terms of the amplitude permeability as 
\begin{align*}
\hat \Psi &=L_a \left( \hat \Psi \right) \hat i \\
\hat i &=\underset{t}{\max} {\left( \hat i_{LF} + i_{HF}(t) \right)}
\end{align*}
and thus it is valid
\begin{align*}
L_a \left( \hat \Psi_{LF} + \frac{\Delta \Psi_{HF}}{2} \right) \hat i &=L_a ^{LF} \hat i_{LF} + \frac{\Delta \Psi_{HF}}{2} \\
L_a ^{LF}&=L_a ^{LF}\left( \hat \Psi_{LF},\Delta \Psi_{HF} \right)
\end{align*}
On that enclosing symmetric major loop, characterized by parameters $\hat \Psi^{MJ}$, $\hat i_{LF}^{MJ}$ and $\hat i^{MJ}$, we have that
\begin{align*}
\hat \Psi^{MJ}&=L_a \left( \hat \Psi^{MJ} \right) \hat i^{MJ}=\hat \Psi \\
\hat i^{MJ}&=\hat i_{LF}^{MJ}=\hat i \\ 
\end{align*}
Consequently, for $\Delta \Psi_{HF}>0$ we get
\begin{align}
L_{\hat{rev}} \left( \hat \Psi^{MJ} \right)&=L_{\hat{rev}} \left( \hat \Psi \right) <L_{\hat{\Delta}} < L_{\hat{rev}} \left( \hat \Psi_{LF} \right) \\
L_{\hat{\Delta}} &=L_{\hat{\Delta}} \left( \hat \Psi_{LF},\Delta \Psi_{HF} \right) \notag
\end{align}

\section{Conclusion} \label{sec:Conclusions}
The objective of this paper is to provide a collection of basic definitions and properties to ground a comprehensive ferrite-core based low-frequency-current biased inductor model for an optimized design method, which is a fundamental tool to properly design and control any type of electronic power converter. The same procedures followed to extract the required parameters of N27 and N87 materials can be easily adapted to obtain that data for other ferrite materials.

\section*{Acknowledgment}

The  first author wants to thank Dr. Hernan Haimovich for his guidance and constructive suggestions.

\end{document}